\documentclass[12pt]{article}
\usepackage[dvips]{graphicx}

\begin{document}
\title{Golden fraction in the theory of
nucleation
}
\author{Victor Kurasov}
\date{ St.Petersburg State University }

\maketitle

\begin{abstract}
The problem of the universal form of the size
spectrum is analyzed. The half-widths of two
wings of spectrum is introduced and it is shown
that their ratio is very close to the golden
fraction. In appendix
it is shown that behind the golden fraction of
an image one can find the information basis,
i.e. the proportion of the golden fraction corresponds to some method
to find extremum. The method to find extrema associated with
Fibonacci numbers also leads to proportions
which can be seen in nature or can be introduced
artificially. The information origin of
proportions is proved theoretically and confirmed
by examples in nature and human life.
\end{abstract}

\section{Universal proportion in the form of the size spectrum}

It is  well known that the phenomenon of a
«golden fraction» is widely spread in nature. This
fact is proven by numerous measurements during
hundreds of years. The most striking feature is
that some fundamental proportions in nature
satisfy the golden fraction.

It is worth seeking the golden fraction in the
process of nucleation. The most natural conditions
are the dynamic ones. Under these conditions
there is a universal form of size spectrum
derived in \cite{TMF}. The form of the universal
spectrum is given by the following formula
$$
f= \exp(x-\exp(x))
$$
in the special coordinates (see \cite{TMF}) after
the special renormalization.

The  spectrum has the amplitude
$$
f_{am} = \exp(-1)
$$
which is attained at $x=0$.

The relaxation length is ordinary defined as the
length where the function is diminished in $\exp(1)$
times. So, here appeared two lengths - one
corresponding to the right wing and that
corresponding to the left wing. We shall denote
them as $-x_1$ and $x_2$. They can be expressed
through the W-Lambert function and have the
following values
$$
x_1 = 1.84
$$
$$
x_2 = 1.14
$$
The ratio
$x_2/x_1$ is very close to the golden fraction
$$
x_2 /x_1 = 0.622
$$
This value is very close to the precise value of
the golden fraction $0.618$.
The relative error is less than
one percent.

The situation is clarified  by fig.1.

\begin{figure}[hgh]

\includegraphics[angle=270,totalheight=8cm]{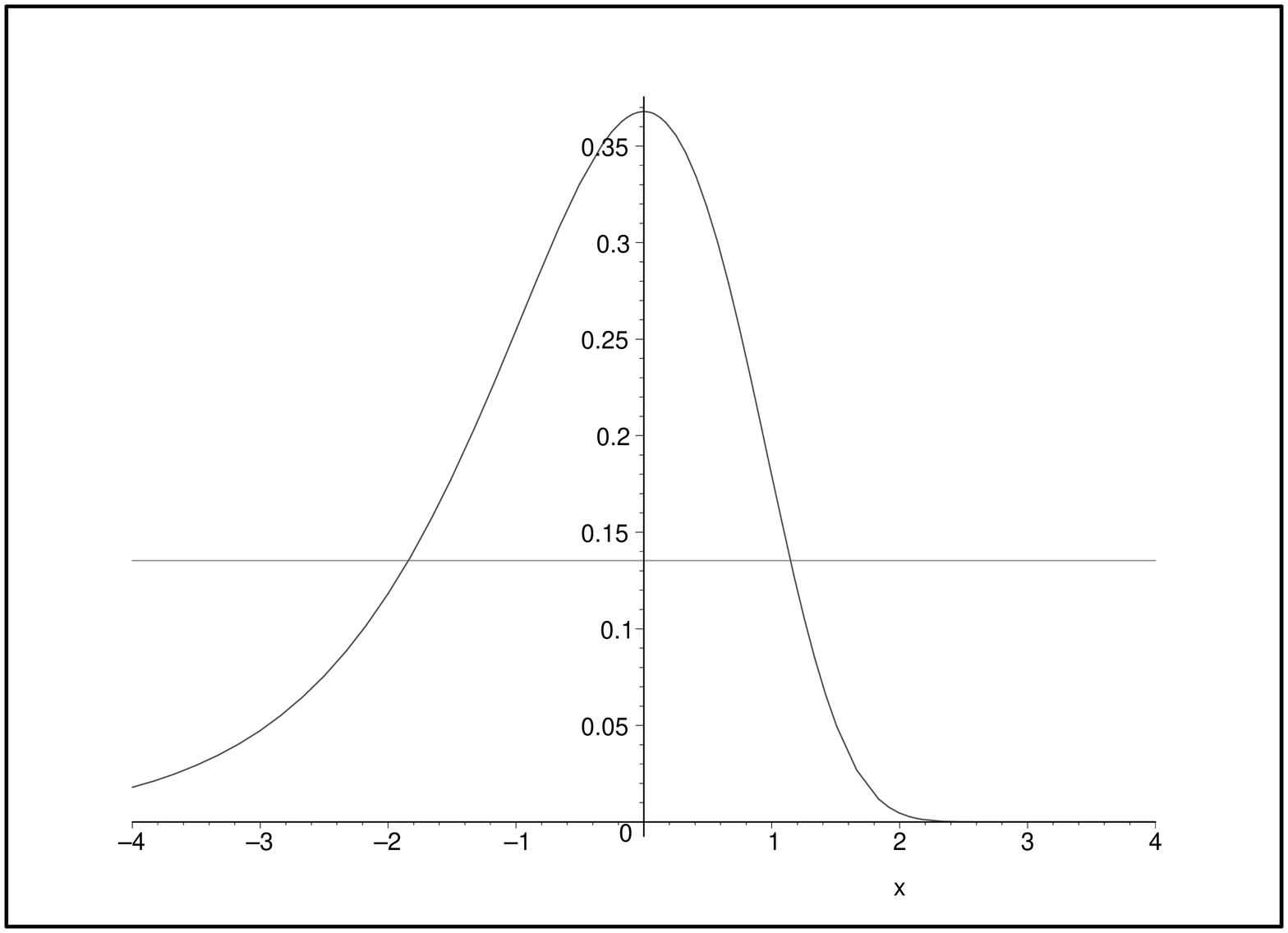}

{ \center  Fig. 1. The form of the universal
spectrum and the golden fraction.
The horizontal line is $e$ times smaller than the amplitude and cuts
the spectrum at the half-widths.}

\end{figure}

There is no yet  any clear interpretation of such
good coincidence of this result with the golden
fraction. It is quite possible
that this is explained by the information origin
of the golden fraction which is derived in
Appendix.

\section{Appendix:The role of the information
interaction in the golden fraction}

The proportions of a
human body satisfy the «golden fraction» rule as it
was stated many times, for example, by Pythagoras, Leonardo da Vinci, etc.
But investigations of Adolf Zeising \cite{Z} showed that only the main
proportion of a male body satisfies in the global
proportions the rule of the «golden fraction». The
global proportion of a female body slightly
differs from the golden fraction $1.618$ and it is
$1.60$. Why this slight deviation takes place?
Below the answer on this question will be given.
This answer is based on the information origin of
the golden fraction which will be analyzed below together with incomplete fractions
appeared as ratios of Fibonacci numbers.

\subsection{Information origin of the golden fraction}

We admit that there is some natural process behind
the phenomena of the golden fraction. What process
can it be? At least this is the process of observation.
Certainly, the process of observation
is the information interaction between the
observer and the environment. What purposes are
attained in this interaction? We suppose that the
observer wants to reconstruct the shape and the
content of the image. The points which produces
the maximal information are certainly the
bifurcation points. The second important class of
points are the points of extrema. Bifurcation
points compose the shape of the object and form
the information background to find all other
characteristics of the object. At this shape the
points of extrema have to be found. So, the
primary task is to find extrema.

It is known that practically all methods to find
exremum different from the simple comparison of
function in different points contain the one
dimensional procedure as a elementary step in the
global procedure \cite{kusin}. So, it is worth to
consider namely the one dimensional procedure of
the extremum seeking.

Consider an elementary interval $[0,1]$. This
interval will be the initial interval where the
extremum (maximum) of the known function exists.
Our task is to determine the position of this
maximum.

To state that there is a maximum inside the given
interval it is necessary to have at least three
points at the interval. Two points will be at the
ends of interval. This is clear because during the
sequential procedure we diminish the initial
interval and automatically the boundary points of a
new interval will be already measured. When we
have three points we can only state that there is
a maximum in the inner point when the function in
the inner point is greater than at the boundary
points. But we can not diminish the interval
without the forth point. Let the inner points be $x_1$
and $x_2$. We have to determine the positions of
these points. Let $x_1$ be less than $x_2$.

When $f(x_1)>f(x_2)$ then we can reduce $[0,1]$ to
$[0,x_2]$. When $f(x_1)<f(x_2)$ then $[0,1]$ can
be reduced to $[x_1,1]$.

The symmetry requires that
$$ x_1=1-x_2$$

Then to determine the position of $x_1$ one can
note that at interval $[0,x_2]$ it will be
necessary to put two points and it would be very
profitable when one of these points coincides with
$x_1$. This point will be the left point in the
interval $[0,x_2]$. Then
$$
x_1=x_2*x_2
$$
or
$$
x_1=(1-x_1)*(1-x_1)
$$
with a root
$$
x_1=\frac{3-\sqrt{5}}{2}
$$
which belongs to $[0,1]$.
The value
$$1-x_1 = \frac{-1+\sqrt{5}}{2}=0.618$$
 is called
the golden fraction.

This value is namely the golden fraction mentioned
at the beginning. So, there appeared a hypothesis
that the golden fraction in nature is associated
with a process of observation and with the
procedure of seeking the extremum. This is the
main idea of this publication. But it is necessary
to confirm this observation. This will be done
below.

\subsection{Example of inapplicability of the pure golden fraction}

We shall check the method of the golden fraction
on example of the seeking for the approximate
extremum by two measurements in the inner points of interval.
After only two measurements it is necessary to take the
final decision.
This is the minimal
number of measurements because one measurement can
not specify the interval smaller than the
initial one.

The simple analysis shows that the smallest
interval will be when two points are $x_1=1/3$,
$x_2=2/3$. It does not correspond to the golden
fraction, but $x_2=0.6666$ is rather close to $0.618$.
Here
\begin{equation}\label{d}
x_1=1-x_2=x_2-x_1
\end{equation}
The reason of the discrepancy is the finite
number of measurements. So, it is necessary to
analyze the optimal procedures to find extremum
with finite number of points.

\subsection{The method to find extremum in the finite number of measurements}

One of the oldest methods to find extrema is the
Fibonacci method described already by Euclid \cite{kusin}.

Let the process be the one dimensional searching
of an extremum restricted by $N$ measurements. The
process is supposed to be a sequential one, i.e.
the observer makes conclusions about the interval
for the possible values of an argument at every
step of the measurements. We shall call this
interval as the uncertainty interval $I_N$.

Now we consider the last measurement $X_N$. It has
to be made in interval $I_{N-1}$. This interval
contains the point of extremum and also the point
$E_{N-1}$ at which the extremum between all
taken measurements is attained.

If we take the new point of measurement $X_N$
equal or very close to $E_{N-1}$ then we get no
new information about the behavior of the function
and such measurement is useless. So, it is
necessary to have a distance between $X_{N}$ and
$E_{N-1}$. Certainly, we do not know this distance
and and speak only about the lowest boundary for
this distance $\delta$.

The best estimate for the $|I_N|$ will be when we
put $X_N$ symmetrically to $E_{N-1}$ with respect
to the middle of interval $I_{N-1}$. Then
$$
I_{N-1} = 2 I_{N} -\delta
$$
This completes the step in the recurrent
procedure.

Now we come to the previous pair of experiments.
The interval $I_{N-1}$ contains $E_{N-2}$. In this
interval two experiments will be made. The best
experiment in this pair will be $E_{N-1}$. Another
experiment will be denoted as $D_{N-1}$. This
point will be the boundary between two parts of
$I_{N-2}$: one part will be included in the
further investigations and the other part will be
thrown out.

But at the beginning of experiment it is not known
what value from the pair will be the best and what
will be thrown out. So, these values have to be
symmetric with respect to the ends of interval.
So, the distances from these points to the ends of
interval have to be equal.

Since both points are symmetric with respect to
the middle of interval and one of the points will
be the optimal $E_{N-1}$ then every point has to
be at the distance $L_{N-1}$ from the end of
interval. Then
$$
L_{N-2} = L_{N-1} + L_{N}
$$
These recurrent relations are typical for the
Fibonacci numbers. It is necessary to check the
initial numbers with $N=1$ and $N=2$ but according
to (\ref{d}) these numbers are equal and after
the renormalization of $L_1$ and $L_2$ we come to
$$ F_1=1, F_2=1$$
Then $L_N=F_N$ are the Fibonacci numbers.

The sequential necessary proportions will be
$$
F_2/F_3=2/3, \ \ F_3/F_4=3/5, \ \ F_4/F_5=5/8
$$
Already $F_4/F_5$ is very close to the golden
fraction and later all sequential fractions will
approach to the golden fraction. So, it is worth to
consider only the first fractions.

One can see that this method is optimal in the
case of $N$ measurements.

\subsection{Examples of proportions}

When human bodies or some other objects in nature
have the mentioned proportions it allows to grasp
their image rather fast. So, one can speak about
the increase of the interaction speed. The time necessary
to get the approximate image is smaller when the
main extreme points of an image coincide with
proportions prescribed by the golden fraction or
Fibonacci fractions.

If our hypothesis is true then there will be
numerous examples of the Fibonacci fractions
$F_2/F_3$ and especially $F_3/F_4$. The higher
fractions can not be observed because they are too
close to the golden fraction. Really, in many cases
it is necessary to get the extremum after two or
three measurements. As an example one can consider
professions of drivers, hunters, etc., where it is
very important to take decisions immediately.

So, one can come to conclusion that there exist
some observers who have the habits to estimate
the extrema in several first steps. The object
under such observation will correspond to their
habits.

As it is known from statistical mechanics the
additional time spending for a fixed job
corresponds to some surplus energy (because the
small time corresponds to the nonequilibrium
process which requires the surplus energy). So,
the construction of the image with ideal
proportions is energetically profitable.

Now it is clear that the proportion of a female
body $0,60$ corresponds to $F_3/F_4$ and it is
explained by historical role of a hunter in a
pre-historical period. Since it was necessary to
take decisions immediately the hunters used to
estimate the image in two or three basic points,
Contrary to men the women have enough time for
observations in their silent life and, thus, the
male body have a proportion of a golden fraction
corresponding to the infinite number of
observations.

Certainly, women can not immediately transform
their body to the golden fraction proportion in
our society where professions of men are now
rather calm. But later the evolution choice will
inevitably bring this proportion to a golden
fraction. The women with long feet are sexually
attractive now and have more chances to get
children, So, earlier or later this proportion
will come to the golden fraction. But it takes
thousands of years and now we have the proportion
$F_3/F_4$ which is the trace of men's professions
in the pre-historical times.

One can see the following interesting example
confirming this theory. The Kuroi in Greece
created before the classic period have proportions
(see fig.2) corresponding to the female fraction
$0.60$.

\begin{figure}[hgh]

\includegraphics[angle=0,totalheight=8cm]{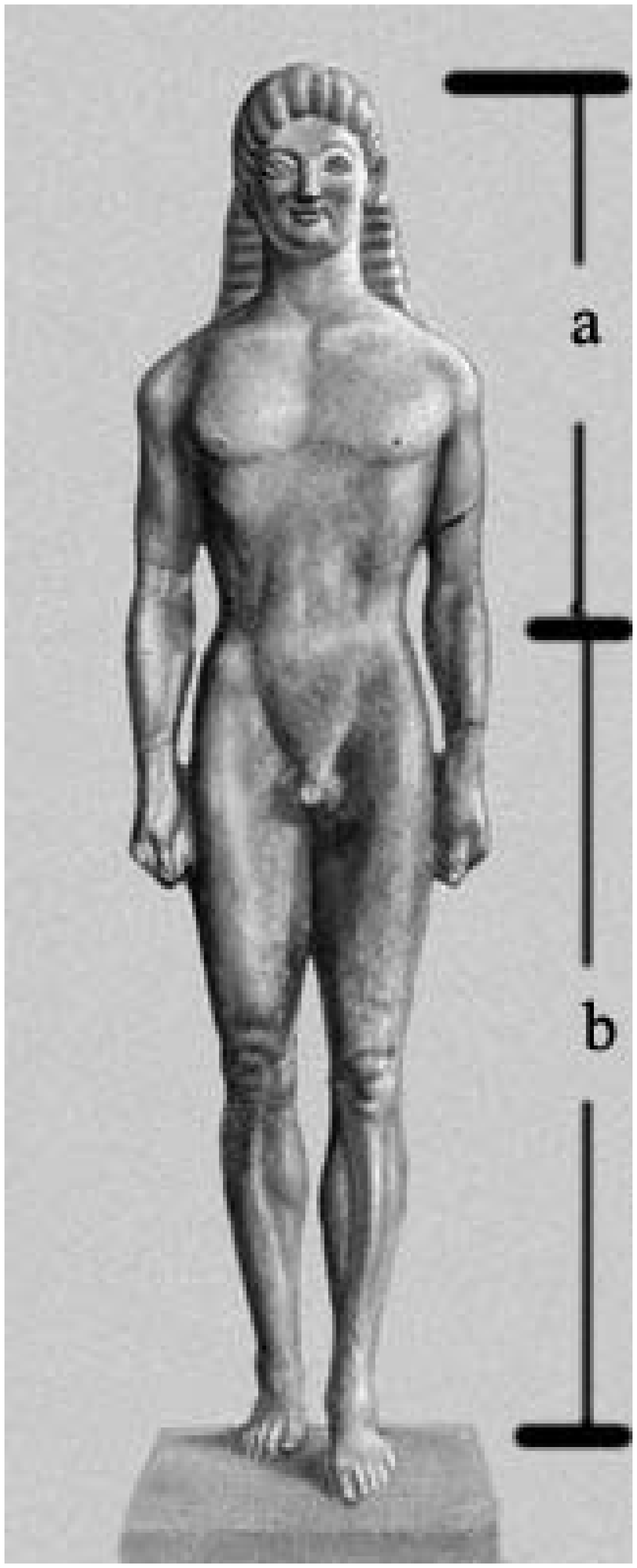}

{ \center Fig. 2. Example of Kuros.
The ratio $b/(a+b)$ is close to $0.60$ }

\end{figure}

An explanation is very simple since the
sculptor and spectators were mainly the men who found the
sexually attractive proportions as the  female ones. Only in classic
period these proportions were reconsidered and brought to the real
proportions of a male body.

Is it possible to view the first proportion
$F_2/F_3$ in a human body?  In nature it does not exist. But it can be seen in
artificial images of women clothes in a fashion
industry images (see fig.3).

\begin{figure}[hgh]

\includegraphics[angle=0,totalheight=8cm]{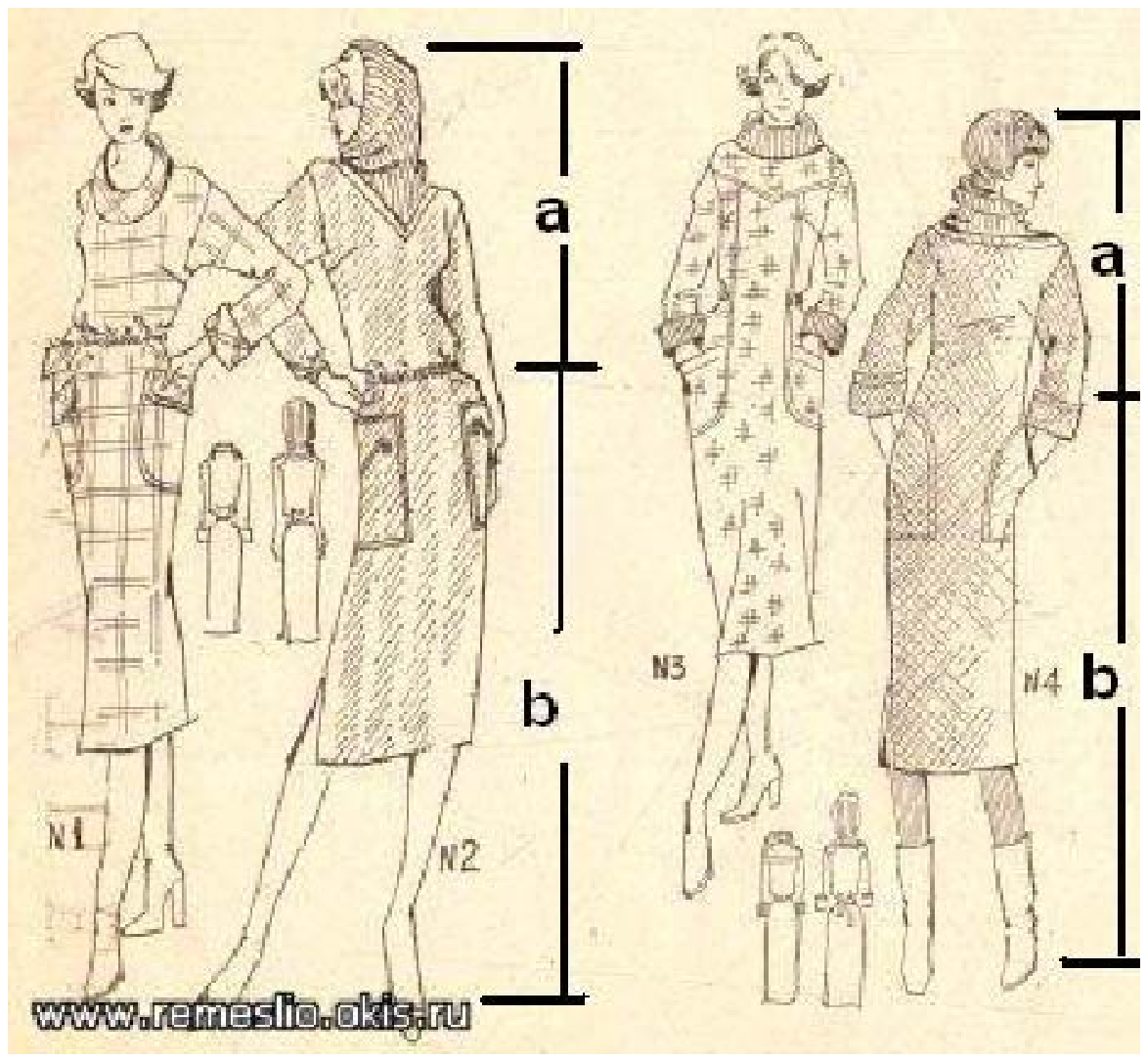}

{ \center Fig. 3. Example of a fashion design.
The ratio $b/(a+b)$ is close to $0.66$ }

\end{figure}

The mentioned main
ratio here  is close to $F_2/F_3$.
One can continue this type of examples. The different heights
of heels help women to modify the main ratio of a
body. One and a half or two centimeters of a heel
give approximately one percent in ratio. So, the
heels in three-four centimeters transforms the ratio
$0.60$ to the golden fraction. This corresponds to
the «English heel». The high heels in 10
centimeters transforms the ratio into the fraction
$F_3/F_4$. This is a «French heel». Two types of
heels give a clear answer on applicability of the
Fibonacci ratios. Women evidently vote by their
heels for the information basis of the harmonic
proportions in nature.

As the result of the given considerations one can
state that now the information origin of
appearance of the golden fraction is clarified.
If we start from the principle of the minimal
energy we can derive the golden fraction
analytically since every observation requires
some time and, thus, some additional energy.
The facts appeared from the incomplete golden
fractions, i.e. from the Fibonacci numbers, show
experimentally that behind the golden proportions
there is the Fibonacci method of the extrema
determination.

One can also mention that now it is clear why  ordinary in
the
women fashion the waist line is
outlined. Really, the waist line goes
approximately three centimeters higher  the
umbilicus point which brings the ratio $0.60=F_3/F_4$ to
the golden fraction.

One can also see that the
spatial sequence of different Fibonacci
proportions introduces the sequence of different
times for observation of these proportions. So,
there appeared
the connection between the space image and the
sequence of times (or the melody) of observation.
This allows to speak about the
space-time connection and about the melody of
paintings.

\end{document}